\documentclass[nofootinbib,amsmath,amssymb,aps,pre,showkeys,onecolumn,eqsecnum,notitlepage]{revtex4-1}
\usepackage{graphicx}  
\usepackage{bm}        

\begin{document}

\title{The Kardar--Parisi--Zhang model of a random kinetic growth: effects of a randomly moving medium}

\author{N.~V.~Antonov$^1$}
\email{n.antonov@spbu.ru}
\author{P.~I.~Kakin$^1$}
\email{polina.kakin@daad-alumni.de}
\author{N.~M.~Lebedev$^{1,2}$}
\email{nikita.m.lebedev@gmail.com}

\affiliation{$^1$ Department of Physics, Saint Petersburg State University,
7/9 Universitetskaya Naberezhnaya, Saint Petersburg 199034, Russia \\
$^2$ N.N. Bogoliubov Laboratory of Theoretical Physics, Joint Institute for Nuclear Research, Dubna 141980, Moscow Region, Russia}

\begin{abstract}
The effects of a randomly moving environment on a randomly growing interface are studied by the field theoretic renormalization group analysis. The kinetic growth of an interface (kinetic roughening) is described by the Kardar--Parisi--Zhang stochastic differential equation while the velocity field of the moving medium is modelled by the Navier-Stokes equation with an external random force. It is found that the large-scale, long-time (infrared) asymptotic behavior of the system is divided into four nonequilibrium universality classes related to the four types of the renormalization group equations fixed points. In addition to the previously established regimes of asymptotic behavior (ordinary diffusion, ordinary kinetic growth process, and passively advected scalar field), a new nontrivial regime is found. The fixed point coordinates, their regions of stability and the critical dimensions related to the critical exponents (e.g. roughness exponent) are calculated to the first order of the expansion in $\varepsilon=2-d$ where $d$ is a space dimension (one-loop approximation) or exactly. The new regime possesses a feature typical to the the Kardar--Parisi--Zhang model: the fixed point corresponding to the regime cannot be reached from a physical starting point. Thus, physical interpretation is elusive. 
\end{abstract}



\maketitle


\section{Introduction}\label{sec:Intro}

An importance of kinetic growth as a study subject takes its root in how wide-spread the phenomenon of a surface kinetic roughening is: it is observed in a flame and smoke propagation, in colloid aggregates' and tumours' growth, in a deposition of a substance on a substrate, and so on~\cite{rost1}--\cite{ball}. As a surface or a phase boundary (interface) evolves with time, it becomes rougher and rougher developing coarser features. In particular, the $n$-th order structure function of a kinetic growth process behaves as~\cite{rost1}--\cite{rost3}:  
\begin{equation}
S_{n}(t,r) \equiv
\langle\left[h(t,{\bf x}) - h(0,{\bf 0})\right]^{n}\rangle \simeq
r^{n\chi}\, F_n (r/t^{1/z}), \quad r=|{\bf x}|.
\label{scaling}
\end{equation}
Here $h(x)=h(t,{\bf x})$ is the height of the interface profile, the
brackets $\langle\dots\rangle$ denote averaging over the statistical
ensemble, $\chi$ and $z$ are the roughness exponent and
the dynamical exponent, respectively, and $F_n(\cdot)$ is a certain
universal scaling function. The power law (\ref{scaling}) describes asymptotic behavior  in the infrared (IR) range (the time $t$ and space $r$ are large in comparison with characteristic microscopic scales). Self-similar (scaling) behavior with universal exponents in the IR range is one of the features of equilibrium nearly-critical systems, thus, universality classes (types of critical or scaling behavior) of kinetic roughening can be established with the methods and approaches developed for the study of critical phenomena. 

While the most typical universality classes of nearly-critical systems
are described by the $\varphi^{4}$-model~\cite{Zinn,Book3}, various microscopic models were proposed to describe kinetic growth: the Eden model~\cite{Eden}, the Edwards--Wilkinson model~\cite{EW},
the restricted solid-on-solid model~\cite{SOS}, the ballistic deposition~\cite{ball}, etc. Among them the Kardar--Parisi--Zhang (KPZ) model~\cite{KPZ} attracts special attention. It is a coarse-grained model of growth described by the nonlinear
stochastic differential equation
\begin{equation}
\partial_{t} h= \varkappa_0\, \boldsymbol{\partial}^{2} h +
\lambda_{0}(\boldsymbol{\partial} h)^{2}/2 + f.
\label{KPZ}
\end{equation}
Here the height field $h(x)=h(t,{\bf x})$ depends on the
$d$-dimensional coordinate ${\bf x}$,
$\partial_{t}= \partial/\partial t$,
$\boldsymbol{\partial}=\{\partial_{i}\}= \{\partial/\partial x_{i}\}$, 
$\boldsymbol{\partial}^{2}=(\boldsymbol{\partial}\cdot\boldsymbol{\partial})=\partial_{i}\partial_{i}$ is the Laplace operator and
$(\boldsymbol{\partial} h)^{2}=(\boldsymbol{\partial} h\cdot\boldsymbol{\partial}h) = \partial_{i}h\partial_{i}h$; the summations over
repeated tensor indices are always implied. The term $\varkappa_0\, \boldsymbol{\partial}^{2} h$ in Eq.~(\ref{KPZ}) describes the surface tension with the
coefficient $\varkappa_0>0$. The term $\lambda_{0}(\boldsymbol{\partial} h)^{2}/2$ models an excess growth along the local normal to the surface. A sign of the parameter $\lambda_{0}$ determines whether the growth is positive or negative. The term $f=f(x)$ is the Gaussian random noise with a zero mean and a pair correlation function
\begin{equation}
\langle f(x)f(x') \rangle = D \,\delta(t-t')
\delta^{(d)}({\bf x}-{\bf x'}),
\label{covar}
\end{equation}
where $D$ is a positive amplitude factor. It is sufficient to consider only the case $D=1$; indeed, any non-trivial amplitude $D$ can be scaled out (absorbed by the fields and other parameters of the model). Thus, we set $D=1$ in the following.

Strictly speaking, a non-vanishing mean value $\langle f\rangle$ should be introduced
to cancel a linear in-time growth of the mean value $\langle h\rangle$. However, quantities like (\ref{scaling}) involve only
differences of the fields, thus, both mean values can be simultaneously
ignored.

As the first two terms on the right-hand side of Eq.~(\ref{KPZ}) are just
the simplest local ones that respect the symmetries $h\to h+$const and
$O(d)$, the KPZ model describes many nonequilibrium, disordered and driven diffusive systems. For example, in~\cite{Uni} the KPZ
model was used in the study of the Universe large-scale matter distribution. Modifications of the KPZ models keep being introduced~\cite{Color}--\cite{Us}.

The field theoretic renormalization group (RG) approach is often used to a great effect in the study of critical phenomena~\cite{Zinn,Book3}. RG approach allows one to find IR attractive fixed points of
renormalizable field theoretic models. The fixed points correspond to universality classes and their critical exponents.

The RG analysis of the KPZ model~\cite{KPZ,FNS,10,11} proved that the field theory related to the stochastic problem
(\ref{KPZ})--(\ref{covar}) is miltiplicatively renormalizable.
The nonlinearity $(\boldsymbol{\partial} h)^{2}$ in Eq.~(\ref{KPZ}) is IR irrelevant
(in the sense of Wilson) for $d>2$, logarithmic (marginal) for $d=2$
and relevant for $d<2$. There is a non-trivial fixed point with the exponents $\chi=0$, $z=2$ but it does not lie in the physical range of the model's
parameters ($\varkappa_{0}>0$; $\lambda_0$ is a real number; $D>0$, if it is introduced in Eq.~(\ref{covar})). All these results are ``perturbatively exact'', i.e., exact in all orders of the expansion
in $\varepsilon\equiv 2-d$. 

The KPZ model, nevertheless, could possess a hypothetical ``essentially non-perturbative''
IR attractive fixed point that is not ``visible'' within any kind of perturbation theory. Under the assumption that the fixed point exists, one can find the exact values for the critical exponents~\cite{KPZ,FNS,Quanta}. As the perturbative solution remains elusive only the functional (``exact'' or ``non-perturbative'') RG is capable of detecting this fixed point~\cite{Canet-aniz,Canet,Canet2}. Some other open problems are discussed in~\cite{Up,Hai}.

Real physical systems become quite sensitive near their critical
points. Indeed, gravity, presence of impurities, external forces, etc., can drastically affect the behavior of the system changing its universality class or even type of a phase transition~\cite{Ivanov}--\cite{Alexa}.

The aim of this paper is to study the influence of the random motion of
the fluid environment on the IR behavior of the randomly growing interface. The advection by the velocity field
$\boldsymbol{v}(x)\equiv \{v_{i}(x)\}$ is introduced by the ``minimal''
replacement
\begin{equation}
\partial_{t} h \to \nabla_{t} h \equiv \partial_{t} h
+ (\boldsymbol{v}\cdot\boldsymbol{\partial}) h = \partial_{t} h
+ v_{i}\partial_{i} h,
\label{nabla}
\end{equation}
where $\nabla_{t}$ is the Galilean covariant (Lagrangean) derivative. The advection is considered to be ``passive'', i.e., the field $h(x)$ is assumed to have no effect on the dynamics of the velocity field $\boldsymbol{v}(x)$. This approximation is sufficient for a preliminary qualitative understanding of what can happen
if the fluid motion is taken into account. Dynamics of the fluid are described by the microscopic model of an incompressible viscous fluid near thermal equilibrium, namely, by the Navier-Stokes equation with a thermal noise as an external random force~\cite{FNS,1,2,27}:
\begin{eqnarray}
\partial_{t} v_{i}
+ (\boldsymbol{v}\cdot\boldsymbol{\partial}) v_{i} = \nu_0 \boldsymbol{\partial}^2 v_{i} - \partial_i P + F_i.
\label{white0}
\end{eqnarray}
Here $\bf v$ is the velocity field vector (it is transverse due to incompressibility: ($\boldsymbol{\partial}\cdot \boldsymbol{v})= \partial_i v_i=0$) with a zero mean, $P$ is the pressure, $\bf F$ is the transverse external random force per unit mass (all of these quantities depend on $x$), $\nu_0$ is the kinematic coefficient of viscosity. The equation (\ref{white0}) is studied on the entire $t$ axis and is supplemented by the retardation condition and by the condition that $\bf v$ vanish asymptotically for $t\rightarrow -\infty$. Random force $\bf F$, a thermal noise, has a Gaussian statistics with a zero mean and a correlation function:
\begin{eqnarray}
\langle F_{i} (t, {\bf x}) F_{j}(t',{\bf x}')\rangle =  \delta(t-t')\,
D_{0}
\int_{k>m} \frac{d{\bf k}}{(2\pi)^{d}} \,
\ P_{ij}({\bf k})\, k^2\,
\exp ({\rm i} ({\bf k}\cdot {\bf r})),
\label{white}
\end{eqnarray}
Here $P_{ij}({\bf k}) = \delta_{ij} - k_i k_j / k^2$ is a transverse 
projector, $k\equiv |{\bf k}|$ is a wave number,
$D_{0}>0$ is an amplitude factor. The integral cutoff at $k=m$, where $m\equiv 1/{\cal L}$ and ${\cal L}$ is an analogue of an integral turbulence scale and provides IR
regularization. Precise form of the cutoff is unimportant, thus, the sharp cutoff is chosen for its simplicity. 

The equation (\ref{white0}) with the random force $\bf F$ (\ref{white}) describes spontaneous velocity fluctuations (relaxation to equilibrium of sufficiently small externally induced fluctuations) and was studied in relation to the problem of long-time tails in Green functions~\cite{FNS}. In the framework of our investigation this choice of the velocity statistics is the most instructive one. Indeed, the both nonlinearities in the scalar equations (\ref{KPZ}) and (\ref{white0}) (the KPZ interaction and the advection term) become logarithmic at $d=2$. This means that they become IR relevant (in the sense of Wilson) simultaneously and, thus, they are equally important for the analysis of the IR asymptotic behavior. If it were not so, one of them would be IR irrelevant for some values of $d$ and would give but corrections to the leading IR asymptotic behavior. 

The paper consists of six Sections. The field theory equivalent to the full stochastic problem (\ref{KPZ}), (\ref{covar}), (\ref{white}) and its diagrammatic technique are described in Section~\ref{sec:Model}. Analysis of the ultraviolet (UV) divergences of the model and discussion of its multiplicative renormalizability are in Section~\ref{sec:Reno}. Derivation of the RG equations is given in Section~\ref{sec:RGE}. Fixed points of the RG equations, possible universality classes and corresponding critical exponents are discussed in Section~\ref{sec:FPS}. It is found out that the RG equations possess a new non-trivial fixed point, in addition to the line of the Gaussian fixed points (trivial regime of critical behavior), purely ``kinematic'' fixed point (the KPZ nonlinearity is irrelevant in the sense of Wilson), and the line of the fixed points related to the pure KPZ model (random motion of the medium is irrelevant). The fixed points' coordinates, their regions of IR stability, and corresponding critical exponents are derived in the leading one-loop order or exactly. 

The new IR attractive fixed point, nevertheless, cannot be reached from a physical starting point. Thus, the model studied in the paper inherits the characteristic feature of the original KPZ model. This problem as well as the obtained results are further discussed in Sec.~\ref{sec:Conc}.

\section{The field theory of the model} \label{sec:Model}

The stochastic problem
(\ref{KPZ}), (\ref{covar}) without the advection is equivalent to the field theory
with the set of fields $\Phi=\{h,h'\}$ and the action functional~\cite{Zinn,Book3}
\begin{equation}
{\cal S}(\Phi)=\frac{1}{2}h'h'+h'\left\{-\partial_{t}h+\varkappa_0
\boldsymbol{\partial}^{2} h+ \frac{1}{2}\lambda_0(\boldsymbol{\partial} h)^{2}\right\}.
\label{act1}
\end{equation}
Here and below, the integrations over $x = (t,{\bf x})$ are always implied, e.g.,
\begin{equation}
\frac{1}{2}h'h'=\frac{1}{2}\int dt\int d{\bf x} \,\, h'(t,{\bf x})\, h'(t,{\bf x}).
\end{equation}

Various correlation functions and response functions of the stochastic problem (\ref{KPZ}), (\ref{covar}) are identified with corresponding Green functions of the field theory (\ref{act1}) (precisely, they are represented by
functional averages over the full set of fields $\Phi=\{ h,h'\}$ with
the weight $\exp {\cal S}(\Phi)$).

The bare propagators are determined by the free part of the action
(\ref{act1}) and have the following form in the frequency--momentum ($\omega$--${\bf k}$) representation:
\begin{eqnarray}
\langle  hh' \rangle_{0} &=& \langle h'h \rangle_{0}^{*}=
\frac{1}{-{\rm i}\omega +\varkappa_0 k^{2}},
\nonumber \\
\langle  hh \rangle_{0} &=&  \frac{1}{\omega^2+\varkappa_0^2 k^4}, \quad
\langle  h'h' \rangle_{0} =  0.
\label{prop1}
\end{eqnarray}
The model includes the interaction vertex $\lambda_0 h'(\boldsymbol{\partial} h)^{2}/2=h'Vhh/2$ with the vertex factor $V=-ik_j (-ip_j)\lambda_0$ where $\boldsymbol{k}$ and $\boldsymbol{p}$ are the momentums flowing out of the vertex via the fields $hh$.

Coupling with the velocity field $\boldsymbol{v}(x)$
is introduced by the substitution (\ref{nabla}) in
Eq.~(\ref{KPZ}) and Eq.~(\ref{act1}). The full problem is then equivalent
to the field theory with the four
fields $\Phi = \{ h,h', \boldsymbol{v}, \boldsymbol{v'}\} $ and the action functional
\begin{eqnarray}
{\cal S}(\Phi)=\frac{1}{2}D_0 \,\partial_i v'_j \partial_i v'_j+v'_l\left\{-\partial_{t}v_l -(\boldsymbol{v}\cdot\boldsymbol{\partial})v_l+
\nu_0 \boldsymbol{\partial}^{2} v_l\right\}+ \frac{1}{2}h'h'+\nonumber\\ 
+h'\left\{-\partial_{t}h-(\boldsymbol{v}\cdot\boldsymbol{\partial})h+
\varkappa_0 \boldsymbol{\partial}^{2} h+
\frac{1}{2}\lambda_0(\boldsymbol{\partial} h)^{2}\right\} 
\label{act2}
\end{eqnarray}
(the summations over
repeated tensor indices are always implied). The transversality of the auxiliary field $\boldsymbol{v'}$ makes it possible to drop the purely longitudinal contribution $\partial_i P$ from Eq.~(\ref{white0}) in Eq.~(\ref{act2}). Since the correlation function (\ref{white}) contains the factor $\boldsymbol{k}^2$ in momentum-frequency representation, the first term of ${\cal S}(\Phi)$ has the factor $\partial_i v'_j\partial_i v'_j$ instead of the usual factor $(\boldsymbol{v'})^2$.

Thus, another four propagators emerge for the full model (\ref{act2}):
\begin{eqnarray}
\langle  \boldsymbol{v}\boldsymbol{v'} \rangle_{0} &=& \langle \boldsymbol{v'}\boldsymbol{v} \rangle_{0}^{*}=
\frac{P_{ij}(\boldsymbol{k})}{-{\rm i}\omega +\nu_0 k^{2}},
\nonumber \\
\langle  \boldsymbol{v}\boldsymbol{v} \rangle_{0} &=&  \frac{D_0 \,k^2 P_{ij}(\boldsymbol{k})}{\omega^2+\nu_0^2 k^4}, \quad
\langle  \boldsymbol{v'}\boldsymbol{v'} \rangle_{0} =  0. 
\end{eqnarray}
There are also the two new vertices: firstly, $v'_l(\boldsymbol{v}\cdot\boldsymbol{\partial})v_l=v'_l V_{l,js} v_j v_s/2$ with the vertex factor $V_{l,js}=i(k_j \delta_{ls} + k_s\delta_{lj})$ where $\boldsymbol{k}$ is the momentum flowing into the vertex via the field $\boldsymbol{v'}$; secondly, $h'(\boldsymbol{v}\cdot\boldsymbol{\partial})h=h'V_j v_j h/2$ with the vertex factor $V_{j}=-ik_j =ip_j$ where $\boldsymbol{k}$ is the momentum flowing into the vertex via the field $h$ and $\boldsymbol{p}$ is the momentum flowing into the vertex via the field $h'$.

There are three coupling constants:
\begin{eqnarray}
g_{0} = D_{0}/\nu_0^3 \sim {\Lambda}^{\varepsilon} , \quad
\widetilde{\lambda_{0}} = \lambda_{0}/\nu_0^{3/2}\sim {\Lambda}^{\varepsilon/2}\\\nonumber
w_0=\varkappa_0/\nu_0.
\label{charges}
\end{eqnarray}
The first two relations are obtained from the dimension analysis
(see the next section) and define the typical UV momentum scale
$\Lambda$. The constant $w_0$ is completely dimensionless and as such must be considered alongside the other coupling constants.

\section{UV divergences and renormalization} \label{sec:Reno}

The analysis of UV divergences is based on the canonical dimensions analysis (``power counting'')~\cite{Zinn,Book3}. There are two independent
scales to be considered in the dynamic models of the type (\ref{act2}): the time scale $T$ and the length scale $L$. The canonical dimension of some quantity $F$ is described by two numbers, the
frequency dimension $d_{F}^{\omega}$ and the momentum dimension $d_{F}^{k}$:
\[[F] \sim [T]^{-d_{F}^{\omega}} [L]^{-d_{F}^{k}}.\]
Normalization conditions
\[ d_k^k=-d_{\bf x}^k=1,\ d_k^{\omega} =d_{\bf x}^{\omega }=0,\
d_{\omega }^k=d_t^k=0,\  d_{\omega }^{\omega }=-d_t^{\omega }=1, \]
and the fact that each term of the action functional is dimensionless further determine $d_{F}^{\omega}$ and $d_{F}^{k}$. The total canonical dimension is defined as $d_{F}=d_{F}^{k}+2d_{F}^{\omega}$
(in the free theory, $\partial_{t}\propto\boldsymbol{\partial}^{2}$). 

Canonical dimensions of the fields and the parameters of the theory (\ref{act2})
are presented in the Table~\ref{table1}.
The table also includes renormalized parameters (the ones without the subscript
``o'') and the renormalization mass $\mu$ that will be introduced later.

\begin{table}[t]
\caption{Canonical dimensions of the fields and the parameters of the theory
(\ref{act2}).}
\label{table1}
\begin{ruledtabular}
\begin{tabular}{c||c|c|c|c|c|c|c|c|c} 
$F$ & $h$ & $h'$ & $\boldsymbol{v}$ & $\boldsymbol{v'}$ & $\varkappa_{0},\varkappa,\nu_0,\nu$ &  $\lambda_{0}^2$ & $g_{0},\widetilde{\lambda_{0}}^2$ &
 $w_{0}, w,g,\widetilde{\lambda}^2$ & $m,\mu,\Lambda$
\\ \hline
$d_{F}^{\omega }$ & $-1/2$ & $1/2$ & $1$ & $-1$ & $1$ & $3$ & $0$ &
$0$ & $0$ 
\\ \hline
$d_{F}^{k}$ & $d/2$ & $d/2$ & $-1$ & $d+1$ & $-2$ & $
-d-4$ &
$2-d\equiv\varepsilon$  & $0$ & $1$ 
\\ \hline
$d_{F}$ & $d/2-1$ & $d/2+1$ & $1$ & $d-1$ & $0$ &
$2-d$ & $\varepsilon$  & $0$ & $1$ 
\\ 
\end{tabular}
\end{ruledtabular}
\end{table}

Thus, the model is logarithmic at $d=2$ when all of the coupling constants
simultaneously become dimensionless. The UV divergences in the Green
functions manifest themselves as poles in $\varepsilon=2-d$.

The total canonical dimension of an arbitrary 1-irreducible Green function
$\Gamma = \langle\Phi \cdots \Phi \rangle_{\rm 1-ir}$ with $\Phi=\{h,h',\boldsymbol{v},\boldsymbol{v'}\}$
in the frequency--momentum representation is given by the relation:
\begin{equation}
d_{\Gamma}=d+2-d_h N_h-d_{h'}N_{h'}-d_v N_v-d_{v'} N_{v'},
\label{dGamma}
\end{equation}
where $N_h,N_{h'},N_v,N_{v'}$ are the numbers of corresponding fields entering
into the function $\Gamma$, see, e.g.,~\cite{Book3}.

The total dimension $d_{\Gamma}$ in the logarithmic theory (i.e., at
$\varepsilon=0$) is the formal index of the UV divergence:
$\delta_{\Gamma}=d_{\Gamma}|_{\varepsilon=0}$. When a number of external momenta occurs as an overall
factor in all diagrams of a certain Green function, the index of
divergence should be adjusted. In the present case the fields $h$ and $\boldsymbol{v'}$ do, indeed, enter the vertices $h'(\boldsymbol{\partial} h)^{2}$, $h'(\boldsymbol{v}\cdot\boldsymbol{\partial})h$ and $v'_l(\boldsymbol{v}\cdot\boldsymbol{\partial})v_l=v'_l v_i \partial_i v_l=-(\partial_i v'_l)v_i v_l$ only in the form of spatial derivatives. Thus, any appearance
of $h$ or $\boldsymbol{v'}$ in some function $\Gamma$ gives an external momentum, and the
real index of divergence is given by the expression
$\delta_{\Gamma}'= \delta_{\Gamma} - N_{h}-N_{v'}$, hence
\begin{equation}
\delta_{\Gamma}'=4 - N_{h} - 2N_{h'} - N_{v} - 2N_{v'}.
\label{IndeX}
\end{equation}

Superficial UV divergences can only be present in 1-irreducible functions that correspond to the non-negative index of divergence $\delta_{\Gamma}'$. 

Canonical dimensions analysis should be augmented by the following considerations. As a manifestation of causality, all the 1-irreducible diagrams without external ``tails'' of the auxiliary fields $\boldsymbol{v'}$, $h'$ involve closed circuits of retarded propagators and, therefore, vanish. Thus, it is sufficient to consider only the functions with $N_{v'}+N_{h'}\ge1$.

The field $h$ is passive in the sense that it does not affect the dynamics of the velocity field. This means that the full Green functions with $N_h=0$ and $N_{h'}>0$ and the 1-irreducible Green functions with $N_h>1$ and $N_{h'}=0$ vanish identically (the numbers $N_{v'}$, $N_{v}$ are arbitrary). In particular, this forbids the divergence in the 1-irreducible function $\langle  v'hh \rangle_{1-ir}$ with the counterterm $v'_i\partial_i h \boldsymbol{\partial}^2 h$. 

The counterterms that have the form of total derivatives (or can be reduced to such form by the integration by parts) vanish after the integration over $x=\{t,\boldsymbol{x}\}$ and should be ignored; consequently, the counterterms that differ by a total derivative should be identified with each other. 

Lastly, the transversality condition $\partial_i v_i=\partial_i v'_i=0$ for the vector fields should not be forgotten.

Taking all of the above into account, one can ascertain that superficial UV divergences can be present only in the following
1-irreducible functions:
\begin{eqnarray}
\langle  h'h' \rangle_{1-ir} \quad (\delta_{\Gamma}=0, \delta_{\Gamma}'=0)
\quad
{\rm with\ the\ counterterm} \quad h'h',
\nonumber \\
\langle h'hh \rangle_{1-ir} \quad (\delta_{\Gamma}=2, \delta_{\Gamma}'=0)
\quad
{\rm with\ the\ counterterm} \quad h'(\boldsymbol{\partial} h)^2,
\nonumber \\
\langle h'h \rangle_{1-ir} \quad (\delta_{\Gamma}=2, \delta_{\Gamma}'=1)
\quad
{\rm with\ the\ counterterm} \quad h'\boldsymbol{\partial}^2 h,
\nonumber \\
\langle \boldsymbol{v}\boldsymbol{v'} \rangle_{1-ir} \quad (\delta_{\Gamma}=2, \delta_{\Gamma}'=1)
\quad
{\rm with\ the\ counterterm} \quad  v'_i\boldsymbol{\partial}^2 v_i,
\nonumber \\
\langle \boldsymbol{v'}\boldsymbol{v}\boldsymbol{v} \rangle_{1-ir} \quad (\delta_{\Gamma}=1, \delta_{\Gamma}'=0)
\quad
{\rm with\ the\ counterterm} \quad  v'_l(\boldsymbol{v}\cdot\boldsymbol{\partial})v_l,
\nonumber \\
\langle \boldsymbol{v'}\boldsymbol{v'} \rangle_{1-ir} \quad (\delta_{\Gamma}=2, \delta_{\Gamma}'=0)
\quad
{\rm with\ the\ counterterm} \quad  \partial_i v'_j\partial_i v'_j,
\nonumber \\
\langle h'h\boldsymbol{v} \rangle_{1-ir} \quad (\delta_{\Gamma}=1,\delta_{\Gamma}'=0)
\quad
{\rm with\ the\ counterterm} \quad  h'(\boldsymbol{v}\cdot\boldsymbol{\partial})h,
\nonumber \\
\langle h'\boldsymbol{v}\boldsymbol{v} \rangle_{1-ir} \quad (\delta_{\Gamma}=0, \delta_{\Gamma}'=0)
\quad
{\rm with\ the\ counterterm} \quad  h'\boldsymbol{v}^{2}.
\end{eqnarray}

Some more additional considerations, related to the symmetry of the 
model, further reduce the number of the counterterms.

The action functional of the KPZ model is invariant with respect to the transformation
\begin{eqnarray}
h(t,{\bf x}) \to h(t,{\bf x}+\boldsymbol{u}t) -
\frac{(\boldsymbol{u}\cdot {\bf x})}{\lambda_0}+\frac{\boldsymbol{u}^2 t}{2\lambda_0}, \quad
h'(t,{\bf x}) \to h'(t,{\bf x}+\boldsymbol{u}t)
\label{Gali1}
\end{eqnarray}
with an arbitrary constant parameter $\boldsymbol{u}$. This invariance is the Galilean symmetry in terms of the vector
field $\partial_{i}h$; it is violated in the full theory (\ref{act1}).
However, the latter possesses another kind of the Galilean symmetry,
namely,
\begin{eqnarray}
h(t,{\bf x}) \to h(t,{\bf x}+\boldsymbol{u}t), \
h'(t,{\bf x}) \to h'(t,{\bf x}+\boldsymbol{u}t) ,
\nonumber \\
\boldsymbol{v} (t,{\bf x}) \to \boldsymbol{v} (t,{\bf x}+\boldsymbol{u}t)
- \boldsymbol{u}, \
\boldsymbol{v'} (t,{\bf x}) \to \boldsymbol{v'} (t,{\bf x}+\boldsymbol{u}t)
\label{Gali2}
\end{eqnarray}
(it is important here that the random force $F$ in the equation describing the velocity field (\ref{white0}) has a factor $\delta(t-t')$ in its correlation function (\ref{white})).
This symmetry puts restrictions on the form of the counterterms, namely, the monomial $h'(\boldsymbol{v}\cdot\boldsymbol{\partial})h$ must enter the
counterterms only in the form of the invariant combination
$h'\nabla_{t}h= h'\partial_{t}h+h'(\boldsymbol{v}\cdot\boldsymbol{\partial} )h$. The first term,
however, is forbidden (the field $h$ must be under the spatial derivative). Thus, the second term is also forbidden. The Galilean symmetry and dimensionality considerations
also rule out the monomials $h'\boldsymbol{v}^{2}$ and $v'_l(\boldsymbol{v}\cdot\boldsymbol{\partial} )v_l$.

All the remaining counterterms ($h'h'$,
$h'\boldsymbol{\partial}^{2}h$, $h'(\boldsymbol{\partial} h)^{2}$,  $v'_i\boldsymbol{\partial}^2 v_i$,  $\partial_i v'_j\partial_i v'_j$) are present
in the action (\ref{act2}). Thus, the theory is multiplicatively renormalizable.
The renormalized action then can be written in the form:
\begin{eqnarray}
{\cal S}_{R}(\Phi)=\frac{1}{2}Z_1 D \,(\partial_i v'_j\partial_i v'_j+v'_l\left\{-\partial_{t}v_l -(\boldsymbol{v}\cdot\boldsymbol{\partial})v_l+
Z_2\nu \boldsymbol{\partial}^{2} v_l\right\}+ \nonumber\\ 
+\frac{1}{2}Z_3 h'h'+h'\left\{-\partial_{t}h-(\boldsymbol{v}\cdot\boldsymbol{\partial})h+
Z_4 \varkappa \boldsymbol{\partial}^{2} h+
\frac{1}{2}Z_5\lambda(\boldsymbol{\partial} h)^{2}\right\}, 
\label{RenAct}
\end{eqnarray}
Here $Z_i$ are the renormalization constants that depend only on the completely
dimensionless parameters $g, w, \widetilde{\lambda}$ and absorb the poles
in $\varepsilon$. The renormalized action ${\cal S}_{R}(\Phi)$ is obtained from the original
one (\ref{act2}) by the renormalization of the fields ($h\rightarrow Z_h h$, $h'\rightarrow Z_{h'} h'$, $\boldsymbol{v}\rightarrow Z_{v} \boldsymbol{v}$, $\boldsymbol{v'}\rightarrow Z_{v'} \boldsymbol{v'}$) and the parameters:
\begin{eqnarray}
\varkappa_{0} = \varkappa Z_{\varkappa}, \quad \nu_{0} = \nu Z_{\nu}, \quad
g_{0} = g \mu^{\varepsilon} Z_{g},      \quad
\widetilde{\lambda}_{0} = \widetilde{\lambda} \mu^{\varepsilon/2} Z_{\widetilde{\lambda}}, \quad w_{0} = w Z_{w}.
\label{Multy}
\end{eqnarray}
The amplitude $D$, the coefficients $\lambda$ and $\varkappa$ are expressed in renormalized parameters as
follows:
\begin{eqnarray}
D = g \nu^3\mu^{\varepsilon}, \quad
\lambda = \nu^{3/2}\widetilde{\lambda}\mu^{\varepsilon/2}, \quad \varkappa=w\nu.
\end{eqnarray}

The renormalization constants in the equations (\ref{RenAct}) and (\ref{Multy})
are subject to the following relations:
\begin{eqnarray}
Z_{g} = Z_{1} \,Z_{2}^{-3}, \quad
Z_{\nu}= Z_{2}, \quad
Z_{w}= Z_{4}\,Z_2^{-1}, \quad
Z_{h} = Z_{3}^{-1/2}, \quad
Z_{h'} = Z_{3}^{1/2}, \nonumber \\
Z_{\widetilde{\lambda}}=Z_5\, Z_3^{1/2}\,Z_2^{-3/2},\quad Z_v=Z_{v'}=1.
\label{ZZ}
\end{eqnarray}

The renormalization constants $Z_{1}$--$Z_{5}$ are calculated directly from
the diagrams, then the constants in Eqs.~(\ref{Multy}) are found from
Eqs.~(\ref{ZZ}). We have performed the calculation to the first order in $g$ and $\widetilde{\lambda}$ (one-loop approximation) using the minimal subtraction (MS) scheme. The details of the calculation are omitted for brevity; a detailed example of a similar calculation see, e.g.,~\cite{Us}. The renormalization constants are as follows:
\begin{eqnarray}
Z_{1} = Z_2=1 - \frac{1}{16\,\varepsilon}\,\hat g, \quad
Z_{3}=1 - \frac{\hat \lambda^2}{w^3}\, \frac{1}{8\,\varepsilon}, \\
Z_{4} = Z_5=1 - \frac{1}{4\,\varepsilon \,w\,(1+w)}\hat g,  \nonumber
\label{Z}
\end{eqnarray}
where $\hat g= g\, S_{d}/(2\pi)^d$, ${\hat \lambda}^2= \widetilde{\lambda}^2\, S_{d}/(2\pi)^d$, and
$S_{d}=2\pi^d/\Gamma(d/2)$ is the area of the unit sphere in $d$
dimensions. The results for $Z_1$ and $Z_2$ agree, up to the notation, with those
obtained in~\cite{FNS}.

\section{RG equations and RG functions} \label{sec:RGE}

The RG equations are written for the renormalized Green functions
$G_{R} =\langle \Phi\cdots\Phi\rangle_{R}$:
\begin{equation}
G(e_{0},\dots) = Z_{h}^{N_{h}} Z_{h'}^{N_{h'}}
G_{R}(e,\mu,\dots).
\label{multi}
\end{equation}
Here $N_{h}$ and
$N_{h'}$ are the numbers of the fields
entering into $G$ (we recall that $Z_{v}=Z_{v'}=1$);
$e_{0}=\{g_{0}, \nu_{0}, w_{0}, \widetilde{\lambda}_0 \}$ is a full set of
bare parameters and $e=\{ g, \nu, w,\widetilde{\lambda}  \}$ are their renormalized
counterparts; the ellipsis stands for the times,
coordinates, momenta, etc.

The differential operation $\widetilde{\cal D}_{\mu}=\mu\partial_{\mu}|_{e_0}$ is expressed in the renormalized variables as follows:
\begin{equation}
{\cal D}_{RG}\equiv {\cal D}_{\mu} + \beta_{g}\partial_{g} +
\beta_{w}\partial_{w}  + \beta_{\widetilde{\lambda}}\partial_{\widetilde{\lambda}} -
\gamma_{\nu}{\cal D}_{\nu},
\label{RG2}
\end{equation}
where ${\cal D}_{x}\equiv x\partial_{x}$ for any variable
$x$. The anomalous dimensions $\gamma$ are defined as
\begin{equation}
\gamma_{F}\equiv \widetilde {\cal D}_{\mu} \ln Z_{F} \quad
{\rm for\ any\ quantity} \ F.
\label{RGF1}
\end{equation}
The $\beta$-functions for the three coupling constants $g$, $w$, and $\widetilde{\lambda}$ are found from the definitions and the
relations (\ref{Multy}):
\begin{eqnarray}
\beta_{g} \equiv \widetilde {\cal D}_{\mu} g = g\,[-\varepsilon-\gamma_{g}],
\quad
\beta_{\widetilde{\lambda}} \equiv \widetilde {\cal D}_{\mu} \widetilde{\lambda} = \widetilde{\lambda}\,[-\varepsilon/2-\gamma_{\widetilde{\lambda}}],\nonumber\\
\beta_{w} \equiv \widetilde {\cal D}_{\mu} w = -w\,\gamma_{w}.
\label{betagw}
\end{eqnarray}

The RG differential equations can be derived by applying the operation $\widetilde{\cal D}_{\mu}$ to the equality (\ref{multi}):
\begin{equation}
\left\{ {\cal D}_{RG} + N_{h} \gamma_{h} +
N_{h'} \gamma_{h'} \right\}
\,G_{R}(e,\mu,\dots) = 0.
\label{RG1}
\end{equation}

At last, equations (\ref{ZZ}) yield the following relations between
the anomalous dimensions (\ref{RGF1}):
\begin{eqnarray}
\gamma_{h}= -\gamma_{3}/2, \quad \gamma_{h'} = \gamma_{3}/2, \quad
\gamma_{w} = \gamma_{4}-\gamma_2, \quad \gamma_{v} = \gamma_{v'}=0,
\nonumber \\
\gamma_{\nu} = \gamma_{2}, \quad
\gamma_{g} = \gamma_{1}-3\gamma_{2}, \quad \gamma_{\widetilde{\lambda}} = \gamma_{5}-3\gamma_2/2+\gamma_3/2.
\label{gadf}
\end{eqnarray}

The anomalous dimension corresponding to a given renormalization constant
$Z_{F}$ can be found from the expression
\begin{equation}
\gamma_{F} = \left(\beta_{g} \partial_{g}+\beta_{w} \partial_{w}+\beta_{\widetilde{\lambda}}\partial_{\widetilde{\lambda}}\right)
\ln Z_{F} \simeq  - \left(\varepsilon {\cal D}_{g}+\varepsilon
{\cal D}_{\widetilde{\lambda}}/2\right) \ln Z_{F},
\label{GfZ}
\end{equation}
obtained from the definition (\ref{RGF1}), expression (\ref{RG2}), and the
fact that the renormalization constants depend only on the three completely
dimensionless coupling constants $g$, $w$, and $\widetilde{\lambda}$. Only
the leading-order terms in the $\beta$-functions (\ref{betagw}) were retained in the second part of
the relation. The MS scheme in the one-loop approximation yields:
\begin{eqnarray}
\gamma_{1} = \gamma_{2}  =\hat{g}/16,
\quad
\gamma_{3} =  \frac{\hat{\lambda}^2}{8w^3}, 
\quad
\gamma_{4} = \gamma_{5}= \frac{\hat g}{4\varepsilon w(1+w)}, 
\label{gammasE}
\end{eqnarray}
where $\hat g$, $\hat{\lambda}$, and $\hat w$ were defined earlier; the corrections of order $\hat{g}^{2}$, $\hat{\lambda}^{4}$ and higher are omitted.

\section{Fixed points, scaling regimes, and critical exponents} \label{sec:FPS}

A long-time large-distance asymptotic behavior
of a renormalizable field theory is determined by IR attractive fixed
points of the RG equations. The coordinates
$g_{*}$, $\widetilde{\lambda}_{*}$, $w_{*}$ of the fixed points of the theory (\ref{act2}) are found from the three equations
\begin{equation}
\beta_{g} (g_{*},\widetilde{\lambda}_{*},w_{*}) = 0, \quad \beta_{\widetilde{\lambda}} (g_{*},\widetilde{\lambda}_{*},w_{*})=0,\quad \beta_{w} (g_{*},\widetilde{\lambda}_{*},w_{*})=0 ,
\label{points}
\end{equation}
with the $\beta$ functions from Eqs.~(\ref{betagw}).
The type of a fixed point is determined by the matrix
\begin{equation}
\Omega=\{\Omega_{ij}=\partial\beta_{i}/\partial g_{j}\},
\label{OmegaDef}
\end{equation}
where $\beta_{i}$ is the full set of the $\beta$ functions and
$g_{j}= \{g,\widetilde{\lambda},w\}$ is the full set of the coupling constants. The real parts of all the $\Omega$ matrix eigenvalues are required to be positive for a fixed point to be IR attractive.

Relations (\ref{betagw}), (\ref{gadf}), and (\ref{gammasE}) yield the explicit
one-loop expressions for the $\beta$ functions:
\begin{eqnarray}
\beta_{g} = g\,[-\varepsilon-\gamma_{g}]=-g\, \left[ \varepsilon + \frac{\hat g}{8}\right], \nonumber \\
\beta_{\widetilde{\lambda}}  = \widetilde{\lambda}\,[-\varepsilon/2-\gamma_{\widetilde{\lambda}}]=-\widetilde{\lambda}\, \left[\frac{\varepsilon}{2}-\frac{3}{32}{\hat g}+\frac{{\hat g}}{4w(w+1)}+\frac{{\hat \lambda}^2}{16w^3}\right] \nonumber \\
\beta_{w} = -w\, \gamma_w=-w\,{\hat g}\, \left[\frac{1}{4w(w+1)}-\frac{1}{16}\right].
\label{betas2}
\end{eqnarray}

The matrix
$\Omega$ turns out to be triangular (because $\partial_w \beta_g=\partial_{\widetilde{\lambda}} \beta_g=\partial_{\widetilde{\lambda}} \beta_w=0$ for any fixed point) and its eigenvalues are given
by the diagonal elements
$\Omega_{g} = \partial \beta_{g} / \partial g$, $\Omega_{{\widetilde{\lambda}}} = \partial \beta_{{\widetilde{\lambda}}} / \partial {\widetilde{\lambda}}$, and
$\Omega_{w} = \partial \beta_{w} / \partial w$.

The fixed points are as follows:

\

\noindent 1. A line of Gaussian (free) fixed points:
$g_{*}=\widetilde{\lambda}_{*}=0$; $w_{*}$ is an arbitrary number; $\Omega_{g} = -\varepsilon$,  $\Omega_{{\widetilde{\lambda}}}=-\varepsilon/2$, $\Omega_{w} = 0$.

\

\noindent 2. Linear passive scalar field fixed point:
$\hat g_{*}=8\varepsilon$; $\widetilde{\lambda}_{*}=0$, $w_{*}=(-1+\sqrt{17})/2$;
$\Omega_{g} = \varepsilon$, $\Omega_{{\widetilde{\lambda}}}=-\varepsilon/4$,  $\Omega_{w} = \varepsilon/2+8\varepsilon/(1+\sqrt{17})^2$.

This fixed point corresponds to the pure linear passive scalar field model, i.e., the KPZ nonlinearity does not affect the leading-order IR asymptotic behavior (it is irrelevant in the sense of Wilson). These results agree, up to the notation and a 
misprint in expression for $w_{*}$, with those
obtained in~\cite{FNS}. It should be noted that due to the different signs before $\varepsilon$ in $\Omega_{g}$ and $\Omega_{{\widetilde{\lambda}}}$ the fixed point never becomes IR attractive.

\

\noindent 3. A curve of the pure KPZ fixed points:
$g_{*}=0$, ${\hat{\lambda}}_{*}^2=-8w_{*}^3\varepsilon$,
$w_{*}$ is an arbitrary number;
$\Omega_{g} = -\varepsilon$, $\Omega_{{\widetilde{\lambda}}}=\varepsilon$, 
$\Omega_{w} = 0$.

This curve of fixed points corresponds to the pure KPZ model, i.e., the random motion of the medium is irrelevant in the sense of Wilson. These fixed points never become IR attractive for the same reasons as the fixed point $2$.

\

\noindent 4.  A new non-trivial fixed point: $\hat g_{*}=8\varepsilon$, $\widetilde{\lambda}_{*}^2=-\varepsilon(-1+\sqrt{17})^3/2$,
$w_{*}=(-1+\sqrt{17})/2$;
$\Omega_{g} = \varepsilon$, $\Omega_{{\widetilde{\lambda}}}=\varepsilon/2$, $\Omega_{w} = \varepsilon/2+8\varepsilon/(1+\sqrt{17})^2$.

This fixed point corresponds to a new nontrivial IR scaling regime
(universality class), in which the nonlinearity of the model
(\ref{act2}) and the random motion of the medium are simultaneously important. However, the point becomes IR attractive when $\varepsilon$ is positive which leads to imaginary $\widetilde{\lambda}_{*}$. Thus, the physical implications of this fixed point are not clear.
The issue will be discussed further in the section~\ref{sec:Conc}.

\

The critical dimension $\Delta_{F}$ of a certain IR relevant quantity $F$ is
given by the relations (see, e.g.,~\cite{Zinn}):
\begin{eqnarray}
\Delta_{F} = d^{k}_{F}+ \Delta_{\omega} d^{\omega}_{F} + \gamma_{F}^{*}
\label{dim}
\end{eqnarray}
(it is assumed that $\Delta_{k} = 1$), where $\Delta_{\omega} = 2 -\gamma_{\nu}^{*}$ is the critical dimension of the frequency, $d^{k,\omega}_{F}$ are the canonical
dimensions of $F$ from the Table~\ref{table1}, and $\gamma_{F}^{*}$ is the
value of the anomalous dimension from Eqs.~(\ref{RGF1}) at the fixed point:
$\gamma_{F}^{*} = \gamma_{F} (g_{*},\widetilde{\lambda}_{*},w_{*})$~\cite{Book3}.

Relations (\ref{gadf}) and explicit one-loop expressions (\ref{gammasE}) yield for the critical dimensions
\[ \Delta_{h} = d/2 - \Delta_{\omega}/2 + \gamma_{h}^{*}, \quad \Delta_{h'} = d/2 + \Delta_{\omega}/2 - \gamma_{h}^{*}, \quad \Delta_{\omega} = 2 -\gamma_{\nu}^{*},  \] 
\[\Delta_{v} = -1 + \Delta_{\omega}, \quad\Delta_{v'} = d+1 - \Delta_{\omega},
\]
where $\gamma_{h} = -{\hat {\lambda}^2}/16 w^3$ and $\quad \gamma_{\nu}={\hat g}/16. $
These relations give the following expressions for the critical dimensions: 
for the line of the fixed points 1:
\begin{eqnarray}
\Delta_{h} = -\frac{\varepsilon}{2}, \quad\Delta_{h'} = 2-\frac{\varepsilon}{2}, \quad \Delta_{\omega} = 2, \quad \Delta_{v} = 1, \quad \Delta_{v'} = 1-\varepsilon,
\label{dimD}
\end{eqnarray}
for the fixed point 2: 
\begin{eqnarray}
\Delta_{h} = -\frac{\varepsilon}{4}, \quad\Delta_{h'} = 2-\frac{3\varepsilon}{4}, \quad \Delta_{\omega} = 2-\frac{\varepsilon}{2}, \quad \Delta_{v} = \Delta_{v'} = 1-\frac{\varepsilon}{2},
\label{dimD2}
\end{eqnarray}
for the curve of the fixed points 3:
\begin{eqnarray}
\Delta_{h} = 0, \quad\Delta_{h'} = 2-\varepsilon,\quad \Delta_{\omega} = 2, \quad \Delta_{v} = 1, \quad \Delta_{v'} = 1-\varepsilon,
\label{dimD3}
\end{eqnarray}
for the fixed point 4:
\begin{eqnarray}
\Delta_{h} = 0, \quad\Delta_{h'} = 2-\varepsilon,\quad \Delta_{\omega} = 2-\frac{\varepsilon}{2}, \quad \Delta_{v} =\Delta_{v'}= 1-\frac{\varepsilon}{2}.
\label{dimD4}
\end{eqnarray}
All of the results for critical dimensions are exact except for the values of $\Delta_{h}$ and $\Delta_{h'}$ in Eqs.~(\ref{dimD4}). Indeed, the critical dimensions for the fixed points 1--3 (and $\Delta_{\omega},\Delta_{v},\Delta_{v'}$) are known exactly due to certain relations between renormalization constants for the case of the passive scalar field~\cite{FNS} (see also~\cite{Book3}) and for the case of the pure KPZ model~\cite{10,11}.

To relate critical dimensions with the critical exponents from (\ref{scaling}) one has to identify $\Delta_{h}= - \chi$ and $\Delta_{\omega}=z$. However, the quantity $S_{n}$ in Eq.~(\ref{scaling}) is not an ordinary $n$-th order Green function of the basic fields $h(x)$ -- it is a sum of pair correlation functions $\langle h^{n-s}(x)h^{s}(0)\rangle$
of the ``composite operators'' $h^{n}(x)$. Renormalization of such quantities usually requires further analysis. However, it turns out that none of the operators $F=h^{n}$ need renormalization and their
critical dimensions are simply given by $\Delta_{F} = n \Delta_{h}$. This
justifies the relation (\ref{scaling}) with the dimensions (\ref{dimD})--(\ref{dimD4}). The proof is omitted here as it is nearly identical to the one reported in~\cite{Us}.

\section{Discussion and conclusion} \label{sec:Conc}

The effects of randomly moving medium on the random kinetic growth of an
interface were studied. The growth was modelled by the Kardar--Parisi--Zhang
stochastic differential equation (\ref{KPZ}), (\ref{covar}).
The random motion of the environment was described by the Navier-Stokes stochastic differential equation with a thermal noise (\ref{white}).

The full problem is equivalent to the
multiplicatively renormalizable theory with the action functional
(\ref{act2}). The field theoretic RG analysis revealed that the system may display four regimes of IR asymptotic behavior related to the four types of possible fixed points
of the RG equations. In addition to previously established universality classes
(ordinary diffusion, ordinary kinetic growth, and passively advected scalar
field), existence of a new nonequilibrium universality class was established.

The fixed point coordinates, their regions of
stability, and corresponding critical dimensions (exponents) were calculated to the first order of the expansion in $\varepsilon=2-d$ (one-loop approximation) or exactly.

However, the predictions of the theory (\ref{act2}) share the same feature with those of the original KPZ model: the coordinates
of the fixed point related to the new universality class lie in the unphysical region $\widetilde{\lambda}_{*}^2<0$. Thus, the physical implications require a careful interpretation. One possible solution may lie in the Doi--Peliti
formalism~\cite{Doi,Tauber}, where the original microscopic problem
is formulated in terms of the creation-annihilation operators. The terms
quadratic in the auxiliary fields can appear in the action functionals
with the negative signs; see e.g.~\cite{Tauber}. The negative term may absorb the ``wrong'' sign of $\widetilde{\lambda}_{*}^2$.

Interestingly, according to the results presented above, when $d=3$ the asymptotic behavior is governed by the trivial Gaussian fixed points; when $d=1$ the fixed point lie in the unphysical region. Nevertheless, kinetic roughening can be observed for the both cases. It seems likely, then, that the IR attractive fixed point of the
pure KPZ model established by the non-perturbative RG~\cite{Canet} might survive
in the current modification of the model. Moreover, new non-perturbative fixed points may also appear. It would be interesting to analyse the model using the non-perturbative RG.

In our analysis we employed the Navier-Stokes equation with a thermal noise. It would be interesting to consider more realistic models with other types of external random force, for example, a non-local noise. This work is in progress.


\section*{Acknowledgments}

The reported study was funded by RFBR according to the research project  18-32-00238. 




\end{document}